\documentclass[pre, twocolumn,showpacs,superscriptaddress,nofootinbibfloatfix]{revtex4-1}

\usepackage{standalone}
\usepackage{graphicx}
\usepackage{color}
\usepackage[usenames,dvipsnames]{xcolor}
\usepackage[colorlinks=true,citecolor=ForestGreen,linkcolor=RubineRed,urlcolor=ForestGreen]{hyperref}
\usepackage{verbatim}
\usepackage{yfonts}
\usepackage{ulem}
\makeatletter

\usepackage{mathptmx} 

\usepackage{amsmath}
\usepackage{amsfonts}
\usepackage{amssymb}
\usepackage{epsfig}
\usepackage{dsfont}
\usepackage{mathrsfs}
\usepackage{enumitem}
\usepackage{graphicx}

\begin{document}

\title{Quantum Work Fluctuations in connection with Jarzynski Equality}
\author{Juan D. Jaramillo}
\affiliation{Department of Physics, National University of Singapore, Singapore 117546}
\author{Jiawen Deng}
\affiliation{NUS Graduate School for Integrative Science and Engineering, Singapore 117597}
\author{Jiangbin Gong} \email{phygj@nus.edu.sg}
\affiliation{Department of Physics, National University of Singapore, Singapore 117546}
\affiliation{NUS Graduate School for Integrative Science and Engineering, Singapore 117597}
\date{\today}

\begin{abstract}
A result of great theoretical and experimental interest, Jarzynski equality predicts a free energy change $\Delta F$ of a system at inverse temperature $\beta$ from an ensemble average of non-equilibrium exponential work, i.e., $\langle e^{-\beta W}\rangle =e^{-\beta\Delta F}$.
The number of experimental work values needed to reach a given accuracy of $\Delta F$ is determined by the variance
of $e^{-\beta W}$, denoted ${\rm var}(e^{-\beta W})$.
 We discover in this work that ${\rm var}(e^{-\beta W})$ in {both harmonic and an-harmonic Hamiltonian systems}
 can systematically diverge in non-adiabatic work protocols, even when the adiabatic protocols do not suffer from such divergence. This divergence may be regarded as a type of dynamically induced phase transition in work fluctuations. For a quantum harmonic oscillator with time-dependent trapping frequency as a working example, any non-adiabatic work protocol is found to yield a diverging ${\rm var}(e^{-\beta W})$ at sufficiently low temperatures, markedly different from the classical behavior. 
 The divergence of
${\rm var}(e^{-\beta W})$ indicates the too-far-from-equilibrium nature of a non-adiabatic work protocol and makes it compulsory
 to apply designed control fields to suppress the quantum work fluctuations in order to test Jarzynski equality.

\end{abstract}
\maketitle







{\it Introduction.} Work fluctuation theorems are a central topic in modern non-equilibrium statistical mechanics \cite{Hanggireview1, Hanggireview2}. One outstanding result is Jarzynski equality (JE) \cite{Jarzynski1} which  makes use of an average of non-equilibrium exponential work, i.e., $e^{-\beta W}$ (where $W$ is the work and $\beta$ is the inverse temperature) to obtain important equilibrium information.  It takes the form $\langle e^{-\beta W}\rangle =e^{-\beta \Delta F}$, where $\langle\cdot\rangle$ represents an ensemble average of all possible work values with the system initially prepared in a Gibbs state, $\Delta F$ is  the free energy difference between initial and final systems at the same $\beta$.  Regarded as a recent breakthrough in non-equilibrium statistical mechanics, JE holds no matter how the work protocol is executed, fast or slow, adiabatic or highly non-adiabatic.  The quantum version of JE  takes precisely the same form \cite{Kurchan1,Tasaki1}, provided that the value of quantum work is interpreted as the energy difference obtained from two energy projective measurements.
 It is also convenient to define the so-called dissipated work  $W_{\rm dis}= W-\Delta F$.  Then  JE gives $\langle e^{-\beta W_{\rm dis}}\rangle=1$.

Early applications of JE focused on biomolecular systems, where it is a well-established method to compute $\Delta F$ \cite{Dellago2}.  Proof-of-principle experiments were carried out \cite{Bustamante1,Kiang1,Ritort1} following an early proposal by Hummer and Szabo \cite{Szabo1}.  Other experiments include those using classical oscillators \cite{Rabbiosi1,Bechinger1}. On the quantum side, the first experiment testing JE was first reported for a quantum spin \cite{Serra1}, relying on interferometric schemes to reconstruct work distributions \cite{Vedral1,Paternostro2}.  While other settings have been proposed, e.g., circuit QED \cite{Campisi1}, a more recent experimental confirmation, following the proposal \cite{Lutz3}, uses a trapped ion subject to an external force \cite{Kihwan1}.  These experiments are by no means straightforward.  In both classical and quantum cases, the ensemble average of a highly nonlinear (in fact, exponential) function of work, as requested by JE, could become practically demanding in yielding a well-converged result for $\Delta F$, as rare events can potentially dominate the average \cite{Jarzynski1,Jarzynski2}.  This in fact has motivated a number of studies in the literature to investigate the errors in predicting $\Delta F$ based on JE  \cite{Zuckerman1,Bustamante2,Zuckerman2,Dellago1,Ritort2,Zuckerman3,Yi1,Kirkpatrick1}.
As suggested by the central limit theorem (CLT), the errors are related to the variance in the exponential work, i.e., ${\rm var}(e^{-\beta W})\equiv \langle e^{-2\beta W}\rangle- e^{-2\beta \Delta F}$.   The larger ${\rm var}(e^{-\beta W})$ is, the more realizations of $W$ we must collect to reach a given accuracy in $\Delta F$.   To obtain $\Delta F$ within an error of $k_BT$, the number of $W$ realizations needed was estimated as ${\rm var}(e^{-\beta W_{\rm dis}})= {\rm var}(e^{-\beta W})/e^{-2\beta \Delta F}$ \cite{Jarzynski3,Dellago1}.


 The quantity  ${\rm var}(e^{-\beta W_{\rm dis}})$ represents the ensemble average of exponential quantities and the Boltzmann exponential cutoff for high-energy states might not always be effective in suppressing the contributions from the high-energy tail.
 Indeed, in contrast to JE itself which is always well behaved, ${\rm var}(e^{-\beta W_{\rm dis}})$ can diverge, suggesting  higher-energy components from the initial Gibbs distribution may contribute even more to the variance.  Surprisingly, prior to our work \cite{Deng1}, this possibility of divergence was rarely mentioned, with one exception \cite{Dellago1} treating an adiabatic work protocol.  According to the principle of minimal work fluctuations (PMWF) \cite{Gaoyang1,Deng1}, once ${\rm var}(e^{-\beta W_{\rm dis}})$ diverges for an adiabatic protocol, then it will suffer from analogous divergence under arbitrary work protocols sharing the same initial and final system Hamiltonians. The accuracy in estimating $\Delta F$  in such situations becomes problematic.   One can no longer rely on the CLT to predict how the error scales with the number of experimental runs.   A generalization of the CLT can prove useful \cite{Deng1} for a specific situation, but in general the error estimate under a diverging ${\rm var}(e^{-\beta W_{\rm dis}})$ is unknown.

 This work discovers the divergence of ${\rm var}(e^{-\beta W_{\rm dis}})$ that is only present for non-adiabatic dynamics.
  Such a possibility is consistent with PMWF. This represents intriguing and more complex situations, where  the ``phase" boundary between a finite ${\rm var}(e^{-\beta W_{\rm dis}})$ and a diverging ${\rm var}(e^{-\beta W_{\rm dis}})$ is yet to be located case by case.
 Our working example below is mainly a quantum harmonic oscillator (QHO) with time-dependent trapping frequency $\omega$ because recent experiments to test quantum JE \cite{Kihwan1} and to construct quantum heat engines \cite{Singer1} were based on QHO ({we consider an-harmonic Hamiltonians in the end}).  Work is done to the system by varying $\omega$.  For a QHO of fundamental and experimental importance, all quantum transition probabilities can be analytically obtained.  
To our great surprise, even in such a prototypic system with all dynamical aspects known for so many years,  divergence of ${\rm var}(e^{-\beta W_{\rm dis}})$ may be induced by non-adiabatic work protocols. The domain of divergence as a function of temperature and other dynamical parameters is found to possess complicated structures.  In general, so long as the dynamics is non-adiabatic, quantum effects at lower temperatures tend to enlarge the domain of divergence in ${\rm var}(e^{-\beta W_{\rm dis}})$. Indeed, even when the classical ${\rm var}(e^{-\beta W_{\rm dis}})$ is well behaved, its quantum counterpart is bound to diverge at sufficiently low temperatures.  This itself constitutes a new aspect of quantum effects in non-equilibrium statistical mechanics. Our results indicate (i) that direct applicability of JE in free energy estimates can become much limited as a system of interest approaches the deep quantum regime, and (ii) that quantum non-adiabatic effects or ``inner friction" \cite{italy1} alone can induce critical changes in work fluctuations. {This work shall also motivate parallel studies on several generalized quantum fluctuation theorems \cite{genFT1,genFT2,genFT3,genFT4,genFT5}.}

{\it  Work characteristic function and model system.} Throughout we focus on work applied to an isolated system.
 Let $H(t)$ be the time-dependent Hamiltonian subject to a work protocol.
The work distribution function is assumed to be $P(W)$.  The Fourier transformation of $P(W)$ yields the so-called work characteristic function \cite{Talkner1}
\begin{equation}
G(\mu)=\int dW\ e^{i\mu W}P(W).
\label{chf-pre}
\end{equation}
For a work protocol starting at $t=0$ and ending at $t=\tau$,  $G(\mu)$ can be expressed as a quantum correlation function $G(\mu)={\rm Tr} \left[e^{iuH_H(\tau)}e^{-iuH(0)}\rho(0)\right]$, where $\rho(0)$ is the initial Gibbs state and $H_H(\tau)$ is the final Hamiltonian in the Heisenberg representation.  While the value of $G(\mu)$ at $\mu=i\beta$ yields JE, the value of $G(\mu)$ at $\mu=2i\beta$ determines  ${\rm var}(e^{-\beta W_{\rm dis}})$.  That is,
\begin{equation}
{\rm var}(e^{-\beta W_{\rm dis}})=\frac{G(2i\beta)}{e^{-2\beta \Delta F}}-1.
\label{eq:var-w-dis}
\end{equation}
This observation indicates that a divergence in $G(2i\beta)$ is equivalent to a divergence in ${\rm var}(e^{-\beta W_{\rm dis}})$ or in ${\rm var}(e^{-\beta W})$.  The quantum correlation function $G(\mu)$ on the imaginary axis is hence our central object. 
An operative expression of \eqref{chf-pre} is given by
%
\begin{equation}
G(\mu)=\sum_{m,n}{\rm exp}\left[i\mu(E_{n}^{\tau}-E_{m}^{0})\right]P_{m,n}^{0\rightarrow \tau}P_{m}^{0},
\label{gmu-app}
\end{equation}
where $P^{0\rightarrow \tau}_{m,n}$ represents the transition probability from $m$th state of $H(0)$ to $n$th state of $H(\tau)$. {Equations (\ref{eq:var-w-dis}) and (\ref{gmu-app}) indicate that, for a system with a finite-dimension Hilbert space such as a spin system,
${\rm var}(e^{-\beta W_{\rm dis}})$ is obtained from a finite summation and will be always well behaved.}

{To lay a solid ground for our surprising findings, we aim to present our main results with computational and analytical calculations supporting each other.  With this in mind, the parametric QHO system seems to be the best choice as a model with an infinite-dimensional Hilbert space.  QHO belongs to the algebraic class $SU(1,1)$ of integrable systems, for which many solutions in the form of exact propagators can be found in the literature \cite{Nikonov,Lohe}.  Systems in the same algebraic class share common dynamical features.  For example, the Calogero-Sutherland model despite being an interacting system is dynamically analogous to QHO \cite{Jaramillo}.  In addition, in the history of quantum mechanics and statistical physics, QHO plays a pivotal role, providing general insights into quantum zero-point energy, low-temperature behavior of the heat capacity of a crystal, open quantum systems, and the relationship between classical mechanics and quantum mechanics, etc.  In particular, in understanding quantum-classical correspondence, tuning the dimensionless ratio of the thermal energy over the energy level spacing is essential to make transitions between classical and quantum regimes.
}

Consider then a QHO with driving in its trapping frequency, with $H_{\text{QHO}}(\omega(t))=\hat{p}^2/2m+m\omega^2(t)\hat{q}^{2}/2$.
Work is done as the trapping frequency changes from $\omega_0=\omega(0)$ to $\omega_1=\omega(\tau)$.
Thanks to Husimi, all $P^{0\rightarrow \tau}_{m,n}$  are available as a function of the Husimi coefficient  $Q^{\ast}\geq 1$ \cite{Husimi1}, {which is defined as
$Q^{\ast}(t)=\frac{1}{2}\left(\omega_0\omega_1X^2+\frac{\omega_0}{\omega_1}\dot{X}^2+\frac{\omega_1}{\omega_0}Y^2+\frac{1}{\omega_0\omega_1}Y^2\right)$, where $X(t)$ and $Y(t)$ are the two solutions of the equation of motion of the corresponding classical  oscillator \cite{Husimi1}}.
For adiabatic dynamics, $Q^{\ast}=1$ and $P^{0\rightarrow \tau}_{m,n}=\delta_{m,n}$.
Any non-adiabaticity is captured by deviations of $Q^{\ast}$ from its adiabatic value $Q^{\ast}=1$.   For later use, we define the compression ratio $\kappa\equiv \omega_0/\omega_1$.  For a sudden quench (sq) protocol $\tau\rightarrow 0$, then $Q^{\ast}\rightarrow Q^{\ast}_{\rm sq}=(\kappa+1/\kappa)/2$.

{\it High-temperature and low-temperature regimes.}
It is of interest to first examine the behavior of $G(\mu=2i\beta)$ in the strictly adiabatic case, i.e., $P^{0\rightarrow\tau}_{m,n}=\delta_{m,n}$.  Because the convergence criteria of the resulting
geometric series in Eq.~(\ref{gmu-app}) becomes $i\mu\omega_1-(i\mu+\beta)\omega_0<0$ with $\mu=2i\beta$, it is obvious that if $\kappa\geq 2$, then $G(2i\beta)$ and hence ${\rm var}(e^{-\beta W_{\rm dis}})$ diverge.  With PMWF, one further concludes that the divergence occurs in all non-adiabatic work protocols.

Next we investigate the behavior of $G(2i\beta)$ under non-adiabatic work protocols, provided that adiabatic work protocols do not
yield divergence in $G(2i\beta)$, i.e., in the regime of $\kappa< 2$.
To that end, we partially resort to a compact expression for $G(\mu)$ \cite{Husimi1}.  That is,
\begin{widetext}
\begin{equation}
G(\mu)=\frac{\sqrt{2}\ (1-e^{-\beta\hbar\omega_0})e^{i\mu\hbar(\omega_1-\omega_0)/2}}{\sqrt{Q^{\ast}(1-e^{2i\mu\hbar\omega_1})(1-e^{-2(i\mu+\beta)\hbar\omega_0})+(1+e^{2i\mu\hbar\omega_1})(1+e^{-2(i\mu+\beta)\hbar\omega_0})-4e^{i\mu\hbar\omega_1} e^{-(i\mu+\beta)\hbar\omega_0}}}.
\label{chf-post}
\end{equation}
\end{widetext}
 Extra care is needed when one extends such a non-analytic
 result based on integration assuming real $\mu$ to the imaginary axis of $\mu$. The Jarzynski equality is always recovered from \eqref{chf-post} at $\mu=i\beta$ \cite{Lutz1}, {\it i.e.},
$G(i\beta)=\frac{\sinh(\beta\hbar\omega_0/2)}{\sinh(\beta\hbar\omega_1/2)}$.
Encouraged by this, we cautiously use this compact expression to help us digest the possible critical boundary of divergence in $G(2i\beta)$ in the regime of $\kappa<2$.  Equation~(\ref{chf-post}) then takes us to the following compact expression for $G(2i\beta)$,
\begin{widetext}
\begin{eqnarray}
G(2i\beta)\rightarrow \frac{\sqrt{2}\sinh(\beta\hbar \omega_0/2)}{\sqrt{\cosh(\beta\hbar(2\omega_1-\omega_0))-1-(Q^{\ast}-1)\ \sinh(2\beta\hbar\omega_1)\sinh(\beta\hbar\omega_0)}}.
\label{2beta}
\end{eqnarray}
\end{widetext}
The real denominator of $G(2i\beta)$ in Eq.~(\ref{2beta}) approaching zero signifies a boundary separating finite values of $G(2i\beta)$ from its diverging values.  We stress however, all such ``shortcut"  conclusions are carefully checked against the original expression in Eq.~(\ref{gmu-app}), with $P^{0\rightarrow \tau}_{m,n}$ in the sum series truncated at some large values of $m$ and $n$ (see Supplementary Material \cite{supp}).

 For the high-temperature (classical) regime $\beta\hbar\omega_{0,1}\ll 1$ we have
\begin{equation}
G(2i\beta) \rightarrow  \frac{\kappa}{\sqrt{(2-\kappa)^2-4\kappa(Q^{\ast}-1)}}.
\label{eq:hot}
\end{equation}
The boundary of divergence in $G(2i\beta)$ is identified at  \begin{equation}
\kappa_c=2\left(Q^{\ast}-\sqrt{Q^{\ast 2}-1}\right).
 \label{eqkappa}
 \end{equation}
Equation ~(\ref{eqkappa}) for adiabatic cases $(Q^{\ast}=1)$ still reproduces the known $\kappa_c=2$ boundary.
In generic non-adiabatic cases with $Q^{\ast}>1$, the divergence boundary is pushed to smaller values, i.e., $\kappa_c<2$, in agreement with the classical result (see Supplementary Material \cite{supp}).    Later on we consider specific work protocols to further digest this result.  One important feature is that
this phase boundary at $\kappa_c$ is no longer dependent upon temperature. Indeed, Eq.~(\ref{eq:hot}) shows that $G(2i\beta)$ itself is temperature-independent in the high-temperature (classical) regime.   Expanding the denominator of Eq.~(\ref{eq:hot}) around $\kappa_c$ under a given $Q^{\ast}$, we find ${\rm var}(e^{-\beta W_{\rm dis}})$ diverges as $\sim (\kappa_c-\kappa)^{-1/2}$ as $\kappa$ approaches the critical value $\kappa_c$ from below.

%
%

%
%
The behavior of $G(2i\beta)$ is entirely different in the low-temperature (deep quantum) regime $\beta\hbar\omega_{0,1}\gg 1$, where
 the compact expression for $G(2i\beta)$ becomes
\begin{equation}
G(2i\beta) \rightarrow  \frac{\exp[\beta\hbar(\omega_0-\omega_1)]}{\sqrt{1-\left(\frac{Q^{\ast}-1}{2}\right)\ \left[\exp(2\beta\hbar\omega_0)-1\right]}}.
\label{eq:cold}
\end{equation}
For strictly adiabatic cases, i.e., $Q^{\ast}=1$, Eq.~(\ref{eq:cold}) is well behaved.  However, for any work protocol with even slight non-adiabaticity, $Q^{\ast}\ne 1$, one observes that the denominator in Eq.~(\ref{eq:cold}) hits zero or becomes imaginary at sufficiently low temperatures. This indicates that ${\rm var}(e^{-\beta W_{\rm dis}})$ must diverge as temperature decreases, so long as the protocol is not strictly adiabatic. The boundary of divergence in ${\rm var}(e^{-\beta W_{\rm dis}})$, within the low-temperature regime, is located at $\beta_c= \ln\left(\frac{Q^{\ast}+1}{Q^{\ast}-1}\right)/(2\hbar\omega_0)$, which does not explicitly depend on $\kappa$.  Further, from Eq.~(\ref{eq:cold}) we find that as $\beta$ approaches $\beta_c$, ${\rm var}(e^{-\beta W_{\rm dis}})$ diverges as
$\sim (\beta_c-\beta)^{-1/2}$.
{To understand the divergence,  one can actually focus on the contribution to ${\rm var}(e^{-\beta W_{\rm dis}})$ made by the transitions from $2n$-th state of $H(0)$ to the ground state of $H(\tau)$. After applying Sterling's
formula to $P^{0\rightarrow \tau}_{2n,0}$ for large $n$, this contribution is found to scale as
$\frac{1}{\sqrt{n}}\left(\frac{Q^{\star}-1}{Q^{\star}+1}\right)^{n}e^{2n\beta\hbar\omega_0}$.
This predicts the same $\beta_c$ as above, because for $\beta>\beta_c$, the more rare the initial state is (sampled from the initial Boltzmann distribution), the more contribution it makes to ${\rm var}(e^{-\beta W_{\rm dis}})$}.  {In other words,
quantum transitions associated with very negative work values, though exponentially suppressed by $\left(\frac{Q^{\star}-1}{Q^{\star}+1}\right)^{n}$ with $n$, still not suppressed as sharply as
in classical cases so as to lose the competition from the exponential increasing factor $e^{2n\beta\hbar\omega_0}$ contained in $e^{-2\beta W}$ at sufficiently low temperatures. This important insight has been confirmed by the behavior of classical and quantum deformed Jarzynzki equalities we just proposed \cite{dJE}.}

\begin{figure}[!htb]
\centering
\includegraphics[width=0.65\linewidth]{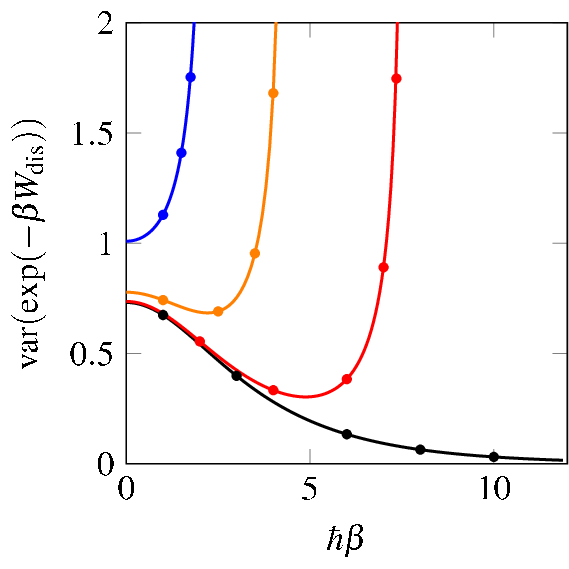}
\caption{${\rm var}(e^ {-\beta W_{\rm dis}})$ vs $\hbar\beta$, with $Q^{\ast}$ chosen to be $1$, $1.01$, $1.1$, $1.5$ (from bottom to top), with $\omega_0=0.35$ and $\omega_1=1$.  The adiabatic case (bottom curve) has no divergence.  Non-adiabatic effects
 induce a blow up in the low temperature regime.  Solid curves are obtained from the
 compact expression \eqref{2beta} and dots from numerics using a  truncation of the quantum numbers in \eqref{gmu-app} with $n_{max}=m_{max}=30$.}
\label{fig_1}
\end{figure}

All these predicted features are checked in numerics, with some examples shown in Fig.~\ref{fig_1}. There the results are obtained based on Eq.~(\ref{gmu-app}) directly as well as from Eq.~(\ref{2beta}). In the high-temperature regime (small $\beta$),  all the plots ${\rm var}(e^{-\beta W_{\rm dis}})$ vs $\hbar\beta$ (in dimensionless units) become flat, in agreement with Eq.~(\ref{eq:hot}). This also indicates that the classical ${\rm var}(e^{-\beta W_{\rm dis}})$ does not diverge in these cases.  For lower temperatures, however,  all the plotted curves tend to blow up, except for the strict adiabatic case (bottom curve) whose ${\rm var}(e^{-\beta W_{\rm dis}})$ approaches zero as temperature decreases, in agreement with Eq.~(\ref{eq:cold}).
 As seen from Fig.~\ref{fig_1}, local minima in ${\rm var}(e^{-\beta W_{\rm dis}})$ as a function of $\beta$ might emerge, reflecting a competition between quantum fluctuations and thermal fluctuations. In addition,  the divergent behavior of ${\rm var}(e^{-\beta W_{\rm dis}})$ close to $\kappa_c$ or $\beta_c$ (obtained from direct numerics and not shown) is also found to agree with the scaling laws of $(\kappa_c-\kappa)^{-1/2}$ and $(\beta_c-\beta)^{-1/2}$.

%

\begin{figure}[!htb]
\centering
\includegraphics[width=0.8\linewidth]{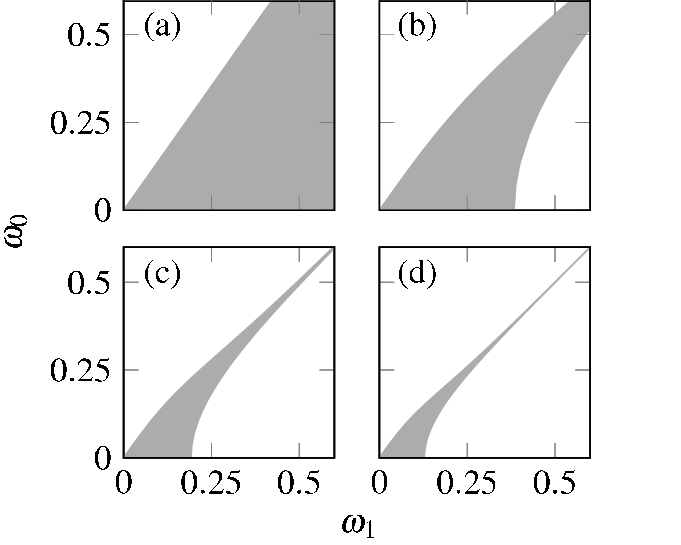}  
\caption{Domain of divergence (white) in ${\rm var}(e^{-\beta W_{\rm dis}})$ for a sudden quench work protocol, with gray areas
indicating finite values of ${\rm var}(e^{-\beta W_{\rm dis}})$.  In the high-temperature regime (a) the domain of divergence is $\omega_{0}\ge\sqrt{2}\omega_1$.  This domain dramatically grows as temperature decreases from (b) to (d), assuming temperatures: $\hbar\beta=5,\ 10,\ 15$.}
\label{fig_2}
\end{figure}

{\it Specific work protocols at intermediate temperatures.} To further investigate the behavior of ${\rm var}(e^ {-\beta W_{\rm dis}})$, we turn to a specific work protocol where the trapping frequency is suddenly quenched from $\omega_0$ to $\omega_1$. In this case $Q^{\ast}=(\kappa+ 1/\kappa)/2$. In the high-temperature regime, Eq.~(\ref{eqkappa}) yields a critical $\kappa_c=\sqrt{2}$, namely, ${\rm var}(e^ {-\beta W_{\rm dis}})$ diverges if  $\omega_0\geq\sqrt{2}\omega_1$.  Fig.~\ref{fig_2} depicts the numerically obtained domain of divergence in ${\rm var}(e^ {-\beta W_{\rm dis}})$ in terms of $\omega_0$ and $\omega_1$, with panel (a) exactly showing this high-temperature behavior.  Quantum effects however dramatically enlarge the domain of divergence (white area) in ${\rm var}(e^ {-\beta W_{\rm dis}})$:  From panel (b) to (d), temperature decreases and  the domain of divergence in ${\rm var}(e^ {-\beta W_{\rm dis}})$ gradually invades the (classical) domain of finite ${\rm var}(e^ {-\beta W_{\rm dis}})$ in panel (a).   For even lower temperatures than shown in Fig.~\ref{fig_2}, the domain of convergence (gray area) collapse into the line $\omega_0=\omega_1$, suggesting any actual quench in $\omega$ will yield divergence in ${\rm var}(e^ {-\beta W_{\rm dis}})$.

 We have also examined a finite-time work protocol, where the parameter $\frac{d{\omega}}{dt}/\omega^{2}(t)$ is chosen to be time-independent, with $\omega(t)=\omega_{1}\omega_0\tau/[\omega_0\tau+t(\omega_1-\omega_0)]$ \cite{delCampo1}.
 Compared with the sudden quench case, the divergence domain is now fragmented into multiple domains with subtle phase boundaries.
In addition, as the temperature decreases, the domain of divergence again grows.
Similar disconnected regions of divergence are observed along the time of driving, where convergent domains are found around the adiabatic times \cite{delCampo1,Uzdin1,Rezek1} for which $Q^{\ast}$ is close to 1. Detailed results are presented in Supplementary Material \cite{supp}.
Finally, it is worth mentioning the possibility of work protocols with $Q^{\ast}>Q^{\ast}_{sq}$, see for example \cite{adam1,Lutz2}. In these cases ${\rm var}(e^{-\beta W_{\rm dis}})$ may even diverge under compression of the harmonic potential ($\kappa<1$) at both high and low temperatures.

\begin{figure}[htbp]
   \centering
\includegraphics[width=3in]{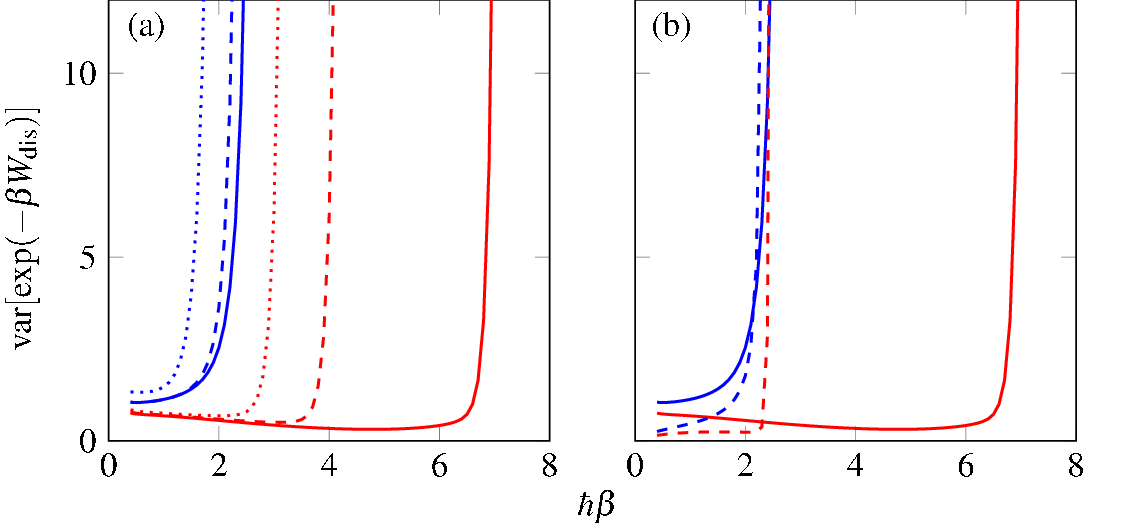}
   \caption{Effect of an additional quartic potential on the work fluctuations of a QHO with $\omega_0=0.35$ and $\omega_1=1$, as described by ${\rm var}[\exp(-\beta W_{\rm dis})]$ vs. $\hbar\beta$.  Panel (a) is for $H_a$ with $J=0$ (solid), 0.01 (dashed) and 0.05 (dotted); the work protocol is of constant compression velocity: $d{\omega}/dt=0.0464$ (red) and $d{\omega}/dt=0.5652$ ( ).  In the absence of the quartic potential term, these sweeping velocities of compression give rise to $Q^{\ast}=1.011$ and $Q^{\ast}=1.5$, respectively.
Panel (b) is for $H_b$, with $J=0$ (solid), 0.01 (dashed); with the same work protocols as in (a). }
   \label{fig:quartic}
\end{figure}

Before ending this section, we also mention our numerical studies of work fluctuations in quantum anharmonic oscillators and other systems such as particle in a box. There we gain similar insights. {In particular, we consider a time-dependent quartic potential
with the total Hamiltonian $H_a(t)=H_{\text{QHO}}(\omega(t))+J\ {\rm cos}(t\pi/\tau) \hat{q}^4$, where $\tau$ is the time of driving, as well as a time-independent quartic potential, with the total Hamiltonian $H_b(t)=H_{\text{QHO}}(\omega(t))+J\ \hat{q}^4$.
There quantum effects are also found to induce the divergence of ${\rm var}(e^{-\beta W_{\rm dis}})$ even if it does not diverge in the classical (high-temperature) domain.
Fig.~\ref{fig:quartic} depicts a few such computational examples.  In all these cases, the effect of a quartic potential is to bring the onset of divergence at higher temperatures as compared with the harmonic cases. }

{\it Conclusions.} JE was hoped to take advantage of non-equilibrium work protocols to estimate the change of free energy in nanoscale and mesoscopic systems. The divergence in ${\rm var}(e^{-\beta W_{\rm dis}})$  presents a hurdle to a direct application of JE.  Note that if the divergence is solely induced by non-adiabatic quantum effects, then one promising solution is to use shortcuts to adiabaticity (STA) \cite{Deng2,delCampo1,Paternostro1,Paternostro1,Muga1}, to go around this divergence and yet still realizing speedy work protocols (but with some price~\cite{Deffner1,Ueda1}).  Indeed, one can even use ${\rm var}(e^{-\beta W_{\rm dis}})$ as a minimization target to design control fields~\cite{Gaoyang1,gaoyang2}.
Because quantum effects are seen to enlarge the domain of divergence drastically in the low temperature regime, using JE for free energy estimates in the deep quantum regime will not be fruitful in the absence of a designed control field.  Nevertheless,  the divergence in ${\rm var}(e^{-\beta W_{\rm dis}})$ offers a new angle to study quantum work fluctuations \cite{Hanggireview2,Quan1,Rahav1} and to characterize the too-far-from-equilibrium nature of a work protocol.  One fascinating question is to study how the divergence of $G(\mu=2i\beta)$ may be reflected in the  behavior of $G(\mu)$ in the real-$\mu$ domain.
Finally, one might wonder if the recent experiment testing quantum JE using a trapped ion \cite{Kihwan1} already suffered from the divergence issue exposed here.  It turns out that this belongs to a fortunate case without any divergence in ${\rm var}(e^{-\beta W_{\rm dis}})$ (see Supplementary Material \cite{supp}). Thus, analyzing possible divergence in exponential work fluctuations will be crucial in guiding the design of future quantum experiments testing JE.

\begin{acknowledgements}
{\bf Acknowledgements:}  This work
is funded by Singapore MOE Academic Research
Fund Tier-2 project (Project No. MOE2014-T2-2- 119 with
WBS No. R-144-000-350-112).  J.G. also acknowledges encouraging discussions with many colleagues.
\end{acknowledgements}


\vspace{2cm}

\appendix

\begin{widetext}
\begin{large} \begin{center} {\bf Appendix } \end{center} \end{large}

As explained in the main text, all divergences in ${\rm var}(e^{-\beta W})$ are traced back to divergences in
\begin{equation}
\langle {\rm e}^{-2\beta W}\rangle=\sum_{m,n}{\rm exp}\left[-2\beta(E_{n}^{\tau}-E_{m}^{0})\right]P_{m,n}^{0\rightarrow \tau}P_{m}^{0},
\label{eq1}
\end{equation}
where $P_{m,n}^{0\rightarrow \tau}$ are the state-to-state transition probabilities and $P_{m}^{0}=e^{-\beta E^{0}_{m}}/Z_0$ is the Gibbs distribution of the initial state.  All our numeric calculations involve a truncation in the series \eqref{eq1}.

%
\section{Transition Probabilities}
%

The transition probabilities between instant eigenstates at $t=0$ and $t=\tau$ for the quantum harmonic oscillator (QHO) with arbitrary driving of the trapping frequency are given by \cite{Husimi1}
\begin{subequations}
\begin{eqnarray}
P^{0\rightarrow\tau}_{2\nu,2\mu}&=&\left(\frac{2}{Q^{\ast}+1}\right)^{1/2}\left(\frac{Q^{\ast}-1}{Q^{\ast}+1}\right)^{\nu+\mu}\frac{(2\mu)!(2\nu)!}{2^{2\mu+2\nu}}\ \left\{\sum_{\lambda=0}^{\min(\nu,\mu)}\frac{\left[-8/(Q^{\ast}-1)\right]^{\lambda}}{(2\lambda)!(\mu-\lambda)!(\nu-\lambda)!}\right\}^2,\\
P^{0\rightarrow\tau}_{2\nu+1,2\mu+1}&=&\left(\frac{2}{Q^{\ast}+1}\right)^{3/2}\left(\frac{Q^{\ast}-1}{Q^{\ast}+1}\right)^{\nu+\mu}\frac{(2\mu+1)!(2\nu+1)!}{2^{2\mu+2\nu}}\ \left\{\sum_{\lambda=0}^{\min(\nu,\mu)}\frac{\left[-8/(Q^{\ast}-1)\right]^{\lambda}}{(2\lambda+1)!(\mu-\lambda)!(\nu-\lambda)!}\right\}^2;
\end{eqnarray}
\label{trans-probs}
\end{subequations}
 where, $\nu,\mu\in\mathbb{N}$, and $Q^{\ast}\geq 1$ is the so called Husimi coefficient.  The selection rules prevent transitions between energy levels with different parity.  
 The definition of $Q^{\ast}$ is \cite{Husimi1}
\begin{equation}
Q^{\ast}(t)=\frac{1}{2}\left(\omega_0\omega_1X^2+\frac{\omega_0}{\omega_1}\dot{X}^2+\frac{\omega_1}{\omega_0}Y^2+\frac{1}{\omega_0\omega_1}Y^2\right),
\label{eq:hc}
\end{equation}
 where $X(t)$ and $Y(t)$ are the two solutions of the equation of motion of the corresponding classical  oscillator,
$\ddot{X}(t)+\omega^2(t)X(t)=0,$
with initial conditions: $X(0)=\dot{Y}(0)=0$ and $\dot{X}(0)=Y(0)=1$.  Any non-adiabaticity is captured by deviations of $Q^{\ast}$ from its adiabatic value, $Q^{\ast}=1$.   The $Q^{\ast}$ used in the main text is assumed to be $Q^{\ast}(\tau)$.

%
\section{Work fluctuations for the classical harmonic oscillator}
%

For classical systems the work distribution is given by
\begin{equation}
P^{c}(W)=\frac{\beta\omega_0}{2\pi}\int dp_0dq_0\ e^{-\beta\mathcal{H}_{0}(p_0,q_0)}\ \delta[W-(\mathcal{H}_\tau(p_0,q_0)-\mathcal{H}_0(p_0,q_0))],
\label{cls:int}
\end{equation}
where $\mathcal{H}_{0}(p_0,q_0)$ is the initial classical Hamiltonian and $\mathcal{H}_{\tau}(p_0,q_0)$ represents the value of the final Hamiltonian $\mathcal{H}_{\tau}$ for the trajectory emanating from $(p_0,q_0)$.
 For the classical harmonic oscillator (CHO) this integral results in two different expressions, associated to the cases: $Q^{\ast}\leq Q^{\ast}_{sq}$ and $Q^{\ast}>Q^{\ast}_{sq}$, where $Q^{\ast}$ is the Husimi coefficient defined in \eqref{eq:hc} and $Q^{\ast}_{sq}$ stands for the Husimi coefficient in sudden quench cases. \\


 We first consider the regime $Q^{\ast}\leq Q^{\ast}_{sq}$, which will be the case if, for example, we have a monotonically increasing or decreasing $\omega(t)$ from $\omega_0$ to $\omega_1$. This gives rise to   positive definite work or negative definite work.   For compression of the harmonic trap $(\omega_0<\omega_1)$ the work distribution reads \cite{Deng2}
\begin{equation}
P^{\hspace{0.05cm}c}_{<}(W)=\beta\ \sqrt{\frac{\kappa}{2(Q^{\ast}_{sq}-Q^{\ast})}}\ \exp\left[\beta\ \frac{\kappa-Q^{\ast}}{2(Q^{\ast}_{sq}-Q^{\ast})}\ W\right]\ I_{0}\left(\beta\ \frac{\sqrt{Q^{\ast 2}-1}}{2(Q_{sq}^{\ast}-Q^{\ast})}\ |W|\right)\Theta(W),
\label{dist-cho}
\end{equation}
where $\kappa\equiv\omega_0/\omega_1$ is the compression ratio, $\Theta(W)$ is the step function and $I_{0}(z)$ is the modified Bessel function of the first kind.  This work distribution coincides with that of the QHO in the semiclassical regime \cite{vanZon1}.
The asymptotic behavior $P^{c}(W\rightarrow\infty)$ determines the convergence of
\begin{equation}
\langle e^{-2\beta W}\rangle_{c}=\int dW\ e^{-2\beta W}P^{c}(W).
\label{cls-2beta}
\end{equation}
 One can then use the approximation: $I_{0}(z)\rightarrow e^{z}/\sqrt{2\pi z},\ z\rightarrow\infty$, to get the tail
\begin{equation}
e^{-2\beta W}P^{c}_{<}(W)\rightarrow\sqrt{\frac{\beta\kappa}{\pi\sqrt{Q^{\ast 2}-1}}}\ \exp\left[\beta\ \left(\frac{\kappa-Q^{\ast}}{2(Q^{\ast}_{sq}-Q^{\ast})}-2+\frac{\sqrt{Q^{\ast 2}-1}}{2(Q_{sq}^{\ast}-Q^{\ast})}\right) W\right].
\label{eq:tail}
\end{equation}
Evaluation of $\langle e^{-2\beta W}\rangle$ requires us to integrate $e^{-2\beta W}P^{c}(W)$ along $W$ from $0$ to $\infty$.  A divergence of $\langle e^{-2\beta W}\rangle_c$ will take place when the exponent in \eqref{eq:tail} is an increasing function of $W$.  Fortunately, the divergence does not show up for compression of the trap. However, for expansion of the trap the work $W$ is always negative, the work function reading
\begin{equation}
P^{\hspace{0.05cm}c}_{<}(W)=\beta\ \sqrt{\frac{\kappa}{2(Q^{\ast}_{sq}-Q^{\ast})}}\ \exp\left[\beta\ \frac{\kappa-Q^{\ast}}{2(Q^{\ast}_{sq}-Q^{\ast})}\ W\right]\ I_{0}\left(\beta\ \frac{\sqrt{Q^{\ast 2}-1}}{2(Q_{sq}^{\ast}-Q^{\ast})}\ |W|\right)\Theta(-W).
\end{equation}
Similarly, the tail is
\begin{equation}
e^{-2\beta W}P^{c}_{<}(W)\rightarrow\sqrt{\frac{\beta\kappa}{\pi\sqrt{Q^{\ast 2}-1}}}\ \exp\left[\beta\ \left(\frac{\kappa-Q^{\ast}}{2(Q^{\ast}_{sq}-Q^{\ast})}-2-\frac{\sqrt{Q^{\ast 2}-1}}{2(Q_{sq}^{\ast}-Q^{\ast})}\right) W\right].
\end{equation}
Since the term $\langle e^{-2\beta W}\rangle$ is an integration along $W$ from $-\infty$ to $0$, the possibility of a negative coefficient, i.e.,
\begin{equation}
\left(\frac{\kappa-Q^{\ast}}{2(Q^{\ast}_{sq}-Q^{\ast})}-2-\frac{\sqrt{Q^{\ast 2}-1}}{2(Q_{sq}^{\ast}-Q^{\ast})}\right)<0
\end{equation}
allows divergence. In particular, for adiabatic expansion of the trap ($Q^{\ast}=1,\ \kappa>1$) the criteria of divergence is $\omega_0>2\omega_1$.\\

Now consider the regime where the nonadiabaticity parameter $Q^{\ast}$ is even larger than that associated with a sudden quench, i.e., $Q^{\ast}>Q^{\ast}_{sq}$. This is possible if we introduce multiple quenches to the trapping frequency $\omega$.  In this case, the integral \eqref{cls:int} for compression or expansion becomes
\begin{equation}
P^{\hspace{0.05cm}c}(W)=\beta\ \sqrt{\frac{\kappa}{2\pi^2(Q^{\ast}-Q^{\ast}_{sq})}}\ \exp\left[-\beta\ \frac{\kappa-Q^{\ast}}{2(Q^{\ast}-Q^{\ast}_{sq})}\ W\right]\ K_{0}\left(\beta\ \frac{\sqrt{Q^{\ast 2}-1}}{2(Q^{\ast}-Q_{sq}^{\ast})}\ |W|\right),
\label{dist-cho-2}
\end{equation}
where $K_{0}(x)$ is the Macdonald function or modified Bessel function of the third kind. 
 This result is consistent with that reported in \cite{Lutz1,vanZon1} using a different approach starting from the characteristic function of the quantum harmonic oscillator (QHO) and taking a semiclassical limit. Similarly, with the approximation $K_{0}(z)\rightarrow \sqrt{\pi}e^{-z}/\sqrt{2 z},\ z\rightarrow\infty$, one obtains the tail
\begin{equation}
e^{-2\beta W}P^{c}(W)\rightarrow\sqrt{\frac{\beta\kappa}{\pi\sqrt{Q^{\ast 2}-1}}}\ \exp\left\{-\beta\ \left[\left(\frac{\kappa-Q^{\ast}}{2(Q^{\ast}-Q^{\ast}_{sq})}+2\right) W+\frac{\sqrt{Q^{\ast 2}-1}}{2(Q^{\ast}-Q^{\ast}_{sq})} |W|\right]\right\}.
\end{equation}
 Because $W$ now can take both positive and negative values,  the tail must go to zero at both $W\rightarrow \pm \infty$ in order to have a finite integral over $W$.  The condition for divergence after integration over $W$ is
\begin{subequations}
\begin{eqnarray}
\frac{\kappa-Q^{\ast}}{2(Q^{\ast}-Q^{\ast}_{sq})}+2+\frac{\sqrt{Q^{\ast2}-1}}{2(Q^{\ast}-Q_{sq}^{\ast})}<0,\ \ \ \ \text{for}\ \ \ W>0;\\
-\frac{\kappa-Q^{\ast}}{2(Q^{\ast}-Q^{\ast}_{sq})}-2+\frac{\sqrt{Q^{\ast2}-1}}{2(Q^{\ast}-Q_{sq}^{\ast})}<0,\ \ \ \ \text{for}\ \ \ W< 0.
\end{eqnarray}
\end{subequations}
Briefly,
\begin{equation}
\left|\frac{\kappa-Q^{\ast}}{2(Q^{\ast}-Q^{\ast}_{sq})}+2\right|>\frac{\sqrt{Q^{\ast2}-1}}{2(Q^{\ast}-Q_{sq}^{\ast})}.
\end{equation}
Combining these results and after a few straightforward steps,  we finally reach a compact condition for divergence
\begin{equation}
\kappa>2(Q^{\ast}-\sqrt{Q^{\ast2}-1}).
\end{equation}

{\it Classical characteristic function.}  First we consider the case $Q^{\ast}<Q^{\ast}_{sq}$.  From the integral formula
\begin{equation}
\int_{0}^{+\infty} dx\ e^{Ax}\ I_{0}(Bx)=-\frac{A}{|A|}\ \frac{1}{\sqrt{A^2-B^2}},\ \ \ |\Re(B)|+\Re(A)\leq 0,
\end{equation}
one obtains from \eqref{cls-2beta} the compact expression
\begin{eqnarray}
\langle e^{-2\beta W}\rangle_{c,<}
=\frac{\sqrt{2\kappa(Q^{\ast}_{sq}-Q^{\ast})}}{\sqrt{[\kappa-Q^{\ast}-4(Q^{\ast}_{sq}-Q^{\ast})]^2-|Q^{\ast 2}-1|}}.
\end{eqnarray}
With some algebra is easy to show that this result coincides with the high-temperature limit of the quantum characteristic function at $\mu=2i\beta$, depicted in Eq.~(7) of the main text.    Next we consider the case $Q^{\ast}>Q^{\ast}_{sq}$.  From the integral
\begin{equation}
\frac{1}{\pi}\int_{-\infty}^{+\infty} dx\ e^{A x}\ K_{0}(B|x|)=\frac{1}{\sqrt{-A^2+B^2}},\ \ \ |\Re(A)|-\Re(B)\leq 0,
\end{equation}
and Eq.~\eqref{dist-cho-2} it is easy to see that $\langle e^{-2\beta W}\rangle_{c,>}$ takes the same form as $\langle e^{-2\beta W}\rangle_{c,<}$, but in the domain $Q^{\ast}>Q^{\ast}_{sq}$.

 %
 \section{Work protocols}
 %

 The existence of additional domains of divergence in ${\rm var}(e^{-\beta W_{\rm dis}})$ away of adiabaticity is allowed by the PMWF.  We study such behavior for two characteristic work protocols: (i) Sudden quench, $Q^{\ast}\rightarrow Q^{\ast}_{sq}$, and (ii) a protocol with  $\frac{d{\omega}}{dt}/\omega^2(t)$ independent of the time of evolution, covering the regime $1<Q^{\ast}\leq Q^{\ast}_{sq}$ for different values of the time of driving $\tau$.  On the other hand, protocols involving multiple quenches \cite{Lutz2,adam1} report $Q^{\ast}$ greater than sudden quench, $Q^{\ast}_{sq}$.  Although not treated here, it is worth mentioning that such protocols with $Q^{\ast}>Q^{\ast}_{sq}$ can lead to divergences in ${\rm var}(e^{-\beta W_{\rm dis}})$ in the high-temperature regime not only for expansion protocols but also for compression.\\

 We briefly recall that
 \begin{equation}
{\rm var}(e^{-\beta W_{\rm dis}})=\frac{G(2i\beta)}{e^{-2\beta \Delta F}}-1,
\label{eq:supp-var-w-dis}
\end{equation}
where $G(\mu)$ is the characteristic function of work.  For the quantum harmonic oscillator
\begin{eqnarray}
G(\mu)=(1-e^{-\beta\hbar\omega_0})e^{i\mu\hbar(\omega_1-\omega_0)/2}\ \left[\sum_{m,n} (e^{i\mu\hbar\omega_1})^{m}(e^{-(i\mu+\beta)\hbar\omega_0})^{n}\ P^{0\rightarrow\tau}_{m,n}\right],
\label{eq:supp-gmu-app}
\end{eqnarray}
where $P^{0\rightarrow\tau}_{m,n}$ are the corresponding transition probabilities.  Our numerical approach is based on truncation of the quantum numbers $n$ and $m$ in the latter expression.  Using the generating function method one arrives at the compact expression (to be derived in the next section)
\begin{equation}
G(\mu)=\frac{\sqrt{2}\ (1-e^{-\beta\hbar\omega_0})e^{i\mu\hbar(\omega_1-\omega_0)/2}}{\sqrt{Q^{\ast}(1-e^{2i\mu\hbar\omega_1})(1-e^{-2(i\mu+\beta)\hbar\omega_0})+(1+e^{2i\mu\hbar\omega_1})(1+e^{-2(i\mu+\beta)\hbar\omega_0})-4e^{i\mu\hbar\omega_1} e^{-(i\mu+\beta)\hbar\omega_0}}}.
\label{eq:supp-chf-post}
\end{equation}
In particular,
\begin{eqnarray}
G(2i\beta)\rightarrow \frac{\sqrt{2}\sinh(\beta\hbar \omega_0/2)}{\sqrt{\cosh(\beta\hbar(2\omega_1-\omega_0))-1-(Q^{\ast}-1)\ \sinh(2\beta\hbar\omega_1)\sinh(\beta\hbar\omega_0)}}.
\label{eq:supp-2beta}
\end{eqnarray}
 For the high-temperature (classical) regime $\beta\hbar\omega_{0,1}\ll 1$, we have
\begin{equation}
G(2i\beta) \rightarrow  \frac{\kappa}{\sqrt{(2-\kappa)^2-4\kappa(Q^{\ast}-1)}},
\label{eq:supp-eq:hot}
\end{equation}
while for the low-temperature (deep quantum) regime $\beta\hbar\omega_{0,1}\gg 1$,
\begin{equation}
G(2i\beta) \rightarrow  \frac{\exp[\beta\hbar(\omega_0-\omega_1)]}{\sqrt{1-\left(\frac{Q^{\ast}-1}{2}\right)\ \left(\exp(2\beta\hbar\omega_0)-1\right)}}.
\label{eq:supp-eq:cold}
\end{equation}
We now discuss the specific work protocols.

 \subsection{Sudden quench}

  For the case of sudden quench, $Q^{\ast}\rightarrow Q^{\ast}_{sq}=(\kappa+1/\kappa)/2$, the domain of divergence in ${\rm var}(e^{-\beta W_{\rm dis}})$ for the high-temperature regime (classical) is $\omega_0>\sqrt{2}\omega_1$, as predicted by Eq.~\eqref{eq:supp-eq:hot}.  As we depart from the high-temperature limit, an extra domain of divergence emerges from the regime of $\omega_1\sim\infty$, reaching smaller values with lower temperatures, see panels (b) to (e) in Fig.~\ref{fig_2suppl}.  This new domain divergence originates from quantum effects as they are absent in the limit of $\hbar\rightarrow 0$.  We can further digest the divergence by looking for the zero and imaginary values of the denominator in the characteristic function in Eq.~\eqref{eq:supp-2beta}, i.e.,
     \begin{equation}
\cosh(\beta\hbar(2\omega_1-\omega_0))-1-(Q^{\ast}_{sq}-1)\ \sinh(2\beta\hbar\omega_1) \sinh(\beta\hbar\omega_0)\leq 0.
\label{eq:supp-cond2}
\end{equation}
 To gain useful insights we consider the limit $\omega_0\rightarrow 0$.  Using  $Q^{\ast}_{sq}\sim 1/2\kappa$,  Eq.~(\ref{eq:supp-cond2}) reduces to
  \begin{equation}
2 \sinh(\beta\hbar\omega_1)-\beta\hbar\omega_1 \cosh(\beta\hbar\omega_1)\leq 0.
\end{equation}
Noting the obvious inequality
\begin{equation}
2 \sinh(\beta\hbar\omega_1)-\beta\hbar\omega_1 \cosh(\beta\hbar\omega_1)<(2-\beta\hbar\omega_1)\ \cosh(\beta\hbar\omega_1),
\end{equation}
the existence of a divergence requires
\begin{equation}
(2-\beta\hbar\omega_1)< 0.
\label{eq:supp-eq:ineq2}
\end{equation}
This divergence is not present in the high-temperature (classical) regime, being associated to $\hbar\beta\rightarrow 0$. Panels (c) to (e)  in Fig.~\ref{fig_2suppl} show such boundary extending to $\omega_0>0$, but always within the compression sector, $\omega_0<\omega_1$.  As an example, the inequality \eqref{eq:supp-eq:ineq2} predicts an onset of divergence around $\omega_1=0.2$ for the temperature $\hbar\beta=10$, in the limit $\omega_0\rightarrow 0$.   This is in agreement with Fig.~\ref{fig_2suppl}(d).
%
%
%
\begin{figure}[!htb]
\centering
\includegraphics{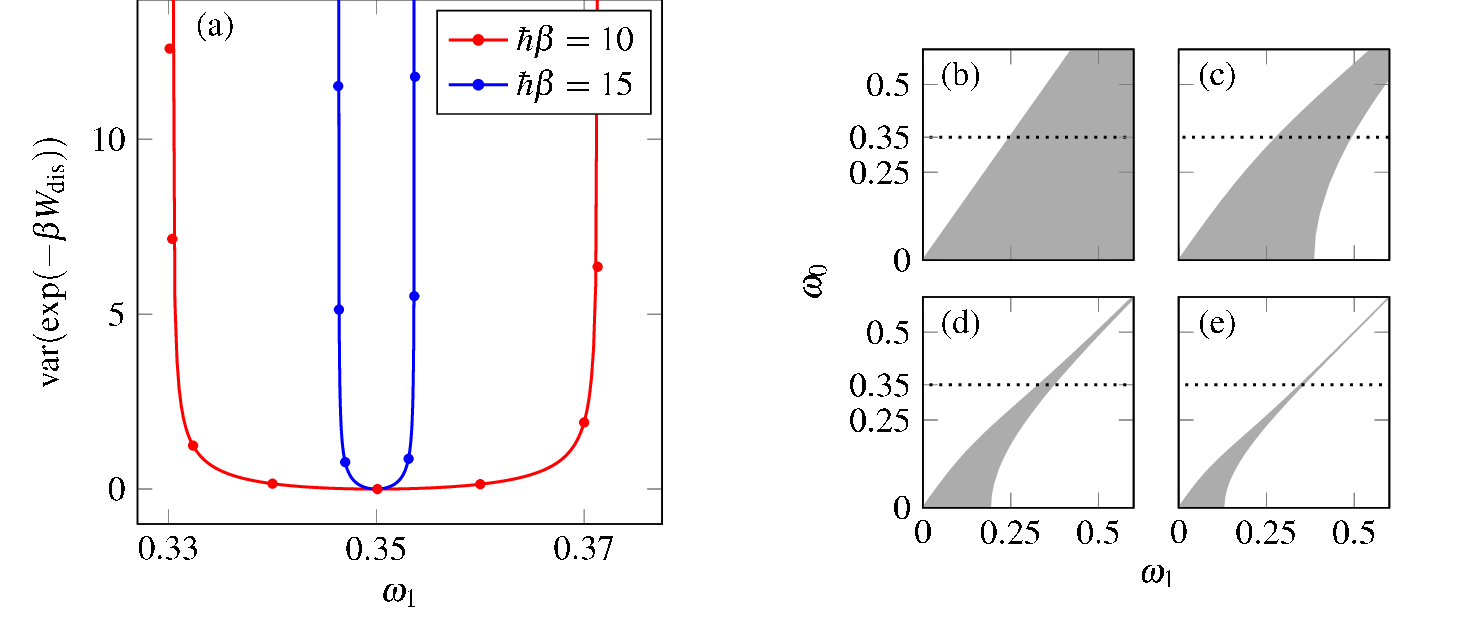}
\caption{Behavior of ${\rm var}(e^{-\beta W_{\rm dis}})$ for sudden quench driving, $Q_{sq}^{\ast}=(\kappa+1/\kappa)/2$.
  The r.h.s.~figures show the domains for divergence/convergence (white/gray area) at temperatures: (b) $\hbar\beta=0$; (c) $\hbar\beta=5$; (d) $\hbar\beta=10$; (e) $\hbar\beta=15$.
    In the high-temperature limit the domain of divergence is, $\omega_{0}\geq\sqrt{2}\omega_1$.  The area of convergence decreases with lower temperatures, eventually collapsing around the  line $\omega_0=\omega_1$.  In the l.h.s.~figure are the profiles for the cross section, $\omega_0=0.35$, at temperatures $\hbar\beta=10,15$; the solid lines are from the compact expression Eq.~\eqref{eq:supp-2beta} and the dots are from numerics based on truncation in Eq.~\eqref{eq:supp-gmu-app} with $n_{max}=m_{max}=100$.}
\label{fig_2suppl}
\end{figure}

\begin{figure}[!htb]
\centering
\includegraphics{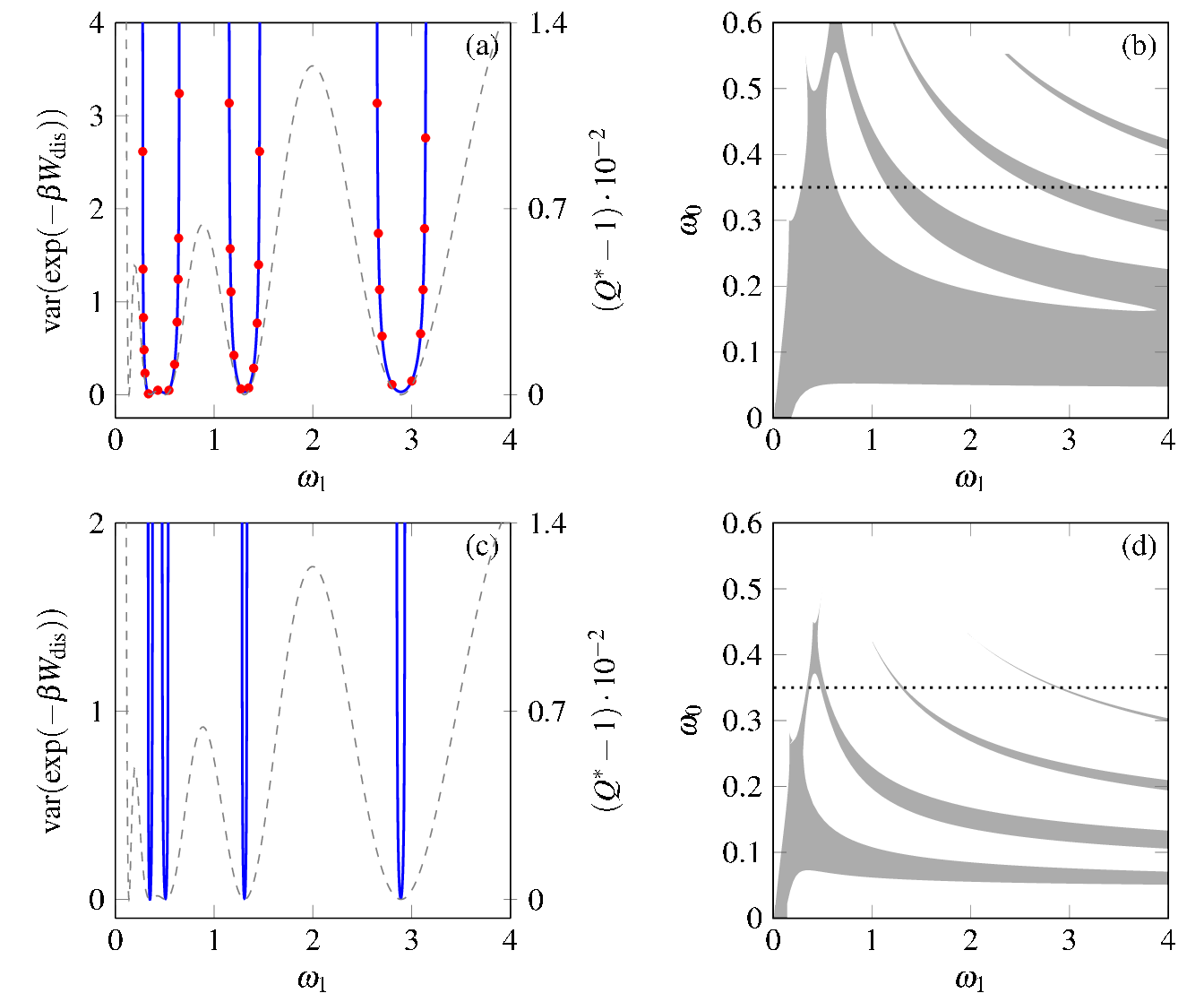}
\caption{Behavior of ${\rm var}(e^{-\beta W_{\rm dis}})$ for the dynamic protocol \eqref{eq:supp-ptcl} with time of driving, $\tau=15$.  We consider the temperature $\hbar\beta=10$ in (a) and (b) and the temperature $\hbar\beta=15$ in (c) and (d).  The panels (b) and (d) show the corresponding domains of divergence/convergence (white/gray area).  The panels (a) and (c) show the profile of ${\rm var}(e^{-\beta W_{\rm dis}})$ and $(Q^{\ast}-1)$ along $\omega_1$, for the cross section $\omega_0=0.35$ which is the dotted line in (b) and (d).   For the panels (a) and (c) the solid   lines depict ${\rm var}(e^{-\beta W_{\rm dis}})$ based on the compact expression Eq.~\eqref{eq:supp-2beta}, while the red dots are from numerics in Eq.~\eqref{eq:supp-gmu-app} based on truncation of the quantum numbers with $n_{max}=m_{max}=50$.  The location of the convergent/divergent behavior is associated to the valleys/peaks of the Husimi parameter, $Q^{\ast}$, as shown in dashed gray of (a) and (c).
}
\label{fig_3suppl}
\end{figure}

\subsection{A protocol with constant $\frac{d{\omega}}{dt}/\omega^2(t)$.}

The second protocol allows us to study the impact of a varying protocol duration $\tau$, between sudden quench ($\tau\rightarrow0$) and the slow driving limit ($\tau\rightarrow\infty$).  The explicit protocol is specified by \cite{delCampo1}
\begin{equation}
\omega(t)=\frac{\omega_{1}\omega_0\tau}{\omega_0\tau+t(\omega_1-\omega_0)}.
\label{eq:supp-ptcl}
\end{equation}
Thanks to a time-independent  $\frac{d{\omega}}{dt}/\omega^2(t)$, one can readily obtain
the corresponding Husimi coefficient
\begin{equation}
Q^{\ast}=\left\{\begin{array}{c} 1+\frac{\cosh(\sqrt{1-\gamma^2})\ln(\kappa)}{1-\gamma^2},\ \text{if}\ \gamma^2\leq 1,\\
\ \ 1+\frac{1}{2}\ \ln(\kappa),\hspace{1cm},\ \text{if}\ \gamma^2=1,\ \\
\ 1+\frac{1-\cos(\sqrt{\gamma^2-1})\ln(\kappa)}{\gamma^2-1},\ \text{if}\ \gamma^2\geq 1;
\end{array}\right.
\label{eq:supp-ptcl-husimi}
\end{equation}
where $\gamma= 2\omega_0\tau/(1-\kappa)$, and $|\frac{d{\omega}}{dt}/\omega^2(t)|=2/\gamma$.  Depicted in Fig.~\ref{fig_3suppl} is the behavior of ${\rm var}(e^{-\beta W_{\rm dis}})$ for a time of driving $\tau=15$ and temperatures $\hbar\beta=10,\ 15$.

%
%

 Compared with the sudden quench case, the divergent domains (see Fig.~\ref{fig_3suppl}(b) and Fig.~\ref{fig_3suppl}(d)) are now fragmented into multiple domains with subtle phase boundaries.  This fragmentation can be partially traced back to the non-monotonic behavior of $Q^{\ast}$ as a function of $\omega_0$ and $\omega_1$ (see dashed lines in Fig.~\ref{fig_3suppl}(a) and Fig.~\ref{fig_3suppl}(c)).  Indeed, for $\left|\frac{\omega}{dt}/\omega^{2}(t)\right|\leq 2$, then $Q^{\ast}=1+\left(1-\cos\left(\gamma^2-1\right)\ln(\kappa)\right)/\left(\gamma^2-1\right)$, where $\gamma=2\omega_0\tau/\left(1-\kappa\right)$.\\

In addition, as the temperature decreases, the domain of divergence tends to grow (the white area in Fig.~\ref{fig_3suppl}(c) is greater than in Fig.~\ref{fig_3suppl}(b)).  In the low temperature limit, the domain for having finite ${\rm var}(e^{-\beta W_{\rm dis}})$ (gray area) shrinks to almost zero area,
leaving only zero-measure boundaries separating different domains of divergence.
Such zero-measure cases with finite ${\rm var}(e^{-\beta W_{\rm dis}})$ arise from  finite-time adiabatic dynamics whose $Q^{\ast}=1$, which is possible because for $|{\omega}/\omega^{2}(t)|\leq 2$, $Q^{\ast}=1$ if  $[1-\cos(\sqrt{\gamma^2-1})\ln(\kappa)]/(\gamma^2-1)=0$.
As mentioned in the main text, similar disconnected regions of divergence are observed along the time of driving, where convergent domains are found around the adiabatic times \cite{delCampo1,Uzdin1,Rezek1}: $\tau_n=\sqrt{1+\frac{4\pi^2n^2}{(\ln\kappa)^2}}|1-\kappa|/(2\omega_0),\ n=1,2,3...$, consistent with the solutions for $Q^{\ast}=1$ in \eqref{eq:supp-ptcl-husimi}.\\

As shown in Fig.~\ref{fig_4suppl}(a), starting from the slow driving limit in a region of $\omega_{0}$ and $\omega_1$ where ${\rm var}(e^{-\beta W_{\rm dis}})$ converges, there is a minimum time of driving $\tau_c$ above which no divergence is found. Below this time there is an irregular alternation of convergent and divergent behavior.  The local minimas in ${\rm var}(e^{-\beta W_{\rm dis}})$  are associated with finite-time adiabatic dynamics.   Note also that for small values of $\omega_0$ similar conclusions as in the sudden quench case can be drawn for the present protocol \eqref{eq:supp-ptcl} because $Q^{\ast}\rightarrow Q_{sq}^{\ast}$ as $\omega_0\rightarrow 0$.\\

\begin{figure}[!htb]
\centering
\includegraphics{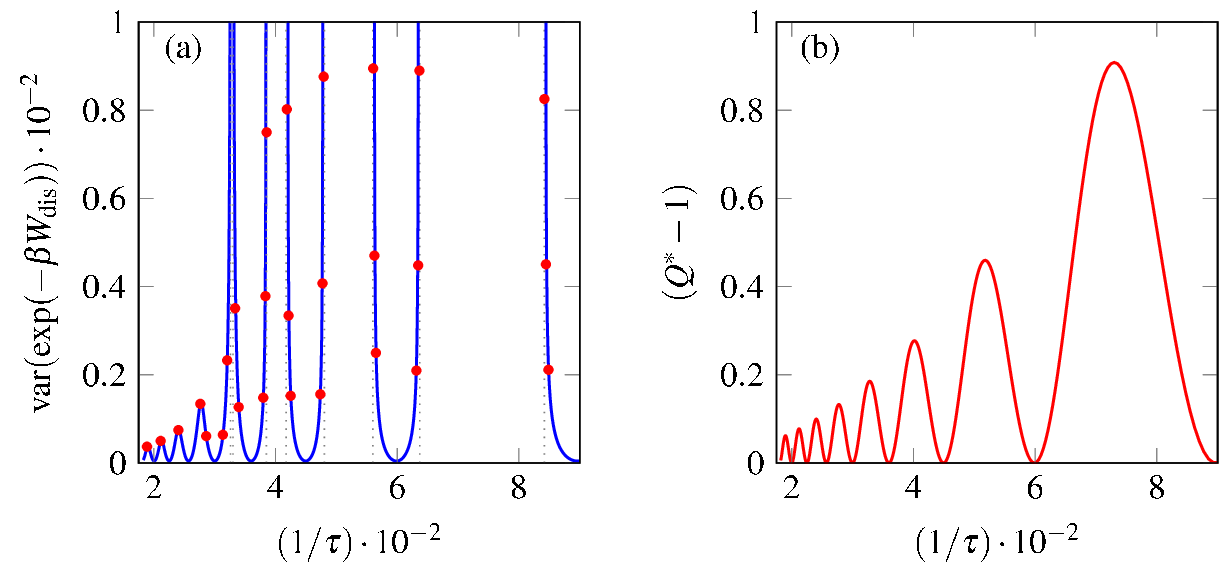}
\caption{Plotted in panel (a) is ${\rm var}(e^{-\beta W_{\rm dis}})$ along the inverse time of driving $1/\tau$, under the work protocol \eqref{eq:supp-ptcl} characterized by a time-independent $\frac{d{\omega}}{dt}/\omega^{2}(t)$.   We fix $\omega_0=0.35, \ \omega_1=1$ and $\hbar\beta=10$.  The solid   curves are based on the compact expression Eq.~\eqref{eq:supp-2beta}, while the red dots are from numerics based on Eq.~\eqref{eq:supp-gmu-app},  with truncation on the quantum numbers: $n_{max}=m_{max}=30$.   The minimas in panel (a) are associated with finite-time adiabatic dynamics \cite{delCampo1}.  Plotted in panel (b) is $(Q^{\ast}-1)$ along the inverse time of driving for the same protocol.}
\label{fig_4suppl}
\end{figure}


%
%

%
\section{Characteristic function for the quantum harmonic oscillator under change of the trap frequency}
%

To compute the characteristic function we use the generating function method \cite{Husimi1}.  The series in Eq.~(3) of the main text has the form $P(u,v)=\sum_{m,n}u^mv^nP_{m,n}^{\tau}$, where $u=e^{i\mu \hbar\omega_1}$ and $v=e^{-(i\mu+\beta) \hbar\omega_0}$.  We review the procedure to arrive at a compact expression for $P(u,v)$.  From the definition of quantum transition probability,
\begin{eqnarray}
P^{0\rightarrow\tau}_{m,n}&=&\left|\int dx_0\int dx (\phi^{\tau}_{m}(x))^{\ast}U(x,x_0;\tau)\phi^{0}_{n}(x_{0})\right|^2,
\end{eqnarray}
where each integral is along the real line.  The series becomes
\begin{eqnarray}
P(u,v)&=&\sum_{m,n=0}^{\infty}u^mv^n\left(\int dx_0\int dx\ (\phi^{\tau}_{m}(x))^{\ast}\ U(x,x_0;\tau)\phi^{0}_{n}(x_{0})\right)\left(\int dy_0\int dy\ (\phi^{\tau}_{m}(y))^{\ast}\ U(y,y_0;\tau)\phi^{0}_{n}(y_0)\right)^{\ast}\nonumber\\
&=&\int\int\int\int dx_0dxdy_0dy \left(\sum_{m=0}^{\infty}u^{m}\phi^{\tau}_{m}(y)(\phi^{\tau}_{m}(x))^{\ast}\right)\ U(x,x_0;\tau)U(y_0,y;-\tau)\ \left(\sum_{n=0}^{\infty}v^n\phi^{0}_{n}(x_{0})(\phi^{0}_{n}(y_{0}))^{\ast}\right),
\label{pre-gen}
\end{eqnarray}
where $U^{\ast}(y,y_0;\tau)=U(y_0,y;-\tau)$.  From Mehler's Hermite polynomial formula, we have
\begin{subequations}
\begin{eqnarray}
\sum_{m=0}^{\infty}u^{m}\phi^{\tau}_{m}(y)(\phi^{\tau}_{m}(x))^{\ast}&=&\sqrt{\frac{m\omega_1}{\hbar\pi(1-u^2)}}\ \exp\left(-\frac{m\omega_1}{2\hbar}\ \frac{(1+u^2)(x^2+y^2)-4uxy}{(1-u^2)}\right),\label{eq-gf-a}\\
\sum_{n=0}^{\infty}v^{n}\phi^{0}_{n}(x_0)(\phi^{0}_{n}(y_0))^{\ast}&=&\sqrt{\frac{m\omega_0}{\hbar\pi(1-v^2)}}\ \exp\left(-\frac{m\omega_0}{2\hbar}\ \frac{(1+v^2)(x_0^2+y_0^2)-4vx_0y_0}{(1-v^2)}\right).
\label{eq-gf-b}
\end{eqnarray}
\label{eq-gf}
\end{subequations}
The latter equations require $\Re(u)^2$ and $\Re(v)^2$ to be lesser than $1$.  Using the well known propagators \cite{Husimi1}
\begin{subequations}
\begin{eqnarray}
U(x,x_0;\tau)&=&\sqrt{\frac{m}{2\pi i\hbar X}}\ \exp\left(\frac{im}{2\hbar X}(\dot{X}x^2-2xx_0+Yx_0^2)\right),\\
U(y_0,y;-\tau)&=&\sqrt{\frac{im}{2\pi\hbar X}}\ \exp\left(-\frac{im}{2\hbar X}(\dot{X}y^2-2yy_0+Yy_0^2)\right);
\end{eqnarray}
\label{ues}
\end{subequations}
one arrives at the Gaussian integral
\begin{eqnarray}
P(u,v)
&=&\sqrt{\frac{m\omega_{1}}{\hbar\pi(1-u^2)}}\ \sqrt{\frac{m\omega_{0}}{\hbar\pi(1-v^2)}}\ \sqrt{\frac{m}{2\pi i\hbar X}}\ \sqrt{\frac{im}{2\pi\hbar X}}\int\int\int\int dx_0dxdy_0dy \exp\left(- {\bf v}^{\rm T}{\bf M}\ {\bf v}\right),
\label{gau}
\end{eqnarray}
where
\begin{equation}
{\bf M}=\frac{m}{2\hbar}\left(\begin{array}{cccc}\frac{1+u^2}{1-u^2}\omega_1 - \frac{i\dot{X}}{X}& \frac{i}{X} & \frac{-2u\omega_1}{1-u^2} & 0 \\ \frac{i}{X} & \frac{1+v^2}{1-v^2}\omega_0 - \frac{iY}{X} & 0 & \frac{-2v\omega_0}{1-v^2} \\ \frac{-2u\omega_1}{1-u^2} & 0 & \frac{1+u^2}{1-u^2}\omega_1+\frac{i\dot{X}}{X} & \frac{-i}{X} \\ 0 & \frac{-2v\omega_0}{1-v^2} & \frac{-i}{X} & \frac{1+v^2}{1-v^2}\omega_0 + \frac{iY}{X}\end{array}\right)\ \ \ \ \text{and}\ \ \ \ {\bf v}=\left(\begin{array}{c} x \\ x_0 \\ y \\ y_0 \end{array}\right).
\end{equation}
Using \eqref{eq:hc} and the identity $\dot{X}Y-X\dot{Y}=1$, one gets
\begin{equation}
{\rm det}({\bf M})=\left(\frac{m}{2\hbar}\right)^4\ 2X^2\omega_0\omega_1\ \left[Q^{\ast}\left(1-u^2\right)\left(1-v^2\right)+\left(1+u^2\right)\left(1+v^2\right)-4uv\right].
\label{detQ}
\end{equation}
From \eqref{gau} it follows that \cite{Husimi1}
\begin{equation}
P(u,v)=\frac{\sqrt{2}}{\sqrt{Q^{\ast}\left(1-u^2\right)\left(1-v^2\right)+\left(1+u^2\right)\left(1+v^2\right)-4uv}}.
\label{puv}
\end{equation}
It is worth noting that numerics supports a broader domain of validity to \eqref{puv} than $\Re(u)^2$ and $\Re(v)^2$ to be lesser than $1$, as required by the generating function method.  This is easy to see in the case of adiabatic dynamics, where $P(u,v)=\sum_{n}(uv)^{n}$ and the convergence condition reduces to $uv<1$, for $u,v\in\mathbb{R}$.

%
\section{Work fluctuations for a quantum harmonic oscillator under an external force}
%

We illustrate how a special work protocol, essentially the one used in the recent proof-of-principle experiment \cite{Kihwan1},  can lead to total absence of divergence in ${\rm var}(e^{-\beta W_{\rm dis}})$.  In the experiment \cite{Kihwan1}, work is done by applying an external force to an ion while keeping the trapping frequency fixed.   The general driving in this case is given by the time-dependent Hamiltonian
\begin{equation}
{H}(t)=\hbar\omega a^{\dag}a+f^{\ast}(t)a+f(t)a^{\dag},
\label{eq:ham-2}
\end{equation}
where $a$ and $a^{\dag}$ are the creation and annihilation operators of a QHO with trapping frequency $\omega$; the complex mechanical parameters are $f$ and $f^{\ast}$, with initial values $f(0)=f^{\ast}(0)=0$.  We note that the qubit in \cite{Kihwan1}, modifying \eqref{eq:ham-2} by a coupling to the external force $(a+a^{\dag})$ via the Pauli matrix $\hat{\sigma}_{x}$, is not initialized at thermal equilibrium but in one of its eigenstates; thus only the phonon is initialized at thermal equilibrium and the qubit is solely used to apply the corresponding work.  The work characteristic function associated with the Hamiltonian in Eq.~\eqref{eq:ham-2} is given by \cite{Talkner2}\\
\begin{equation}
{G}(\mu)=\exp\left(\frac{i\mu |f(\tau)|^2}{\hbar\omega}\right)\ \exp\left[(e^{i\mu\hbar\omega}-1)|z|^2\right]\ \sum_{n=0}^{\infty}P_{n}^{0}L_{n}\left(4|z|^2\sin^2(\hbar\omega\mu/2)\right),
\label{eq:cf-2}
\end{equation}
where $L_n(x)$ are Laguerre polynomials and again we assume an initial state at thermal equilibrium, $P^{0}_{n}=e^{-\beta E^{0}_{n}}/Z_0$.  The non-adiabatic dynamics is captured by the {\it rapidity parameter}
\begin{equation}
z(\tau)=\frac{1}{\hbar\omega}\int_{0}^{\tau}dt \dot{f}(t)\exp(i\omega t).
\end{equation}
 Using again $\langle e^{-2\beta {W}}\rangle={G}(2i\beta)$, one finds that ${G}(2i\beta)$ is always finite.
 Indeed,
from Perron's formula 
\begin{equation}
L_{n}(x)=2^{-1}\pi^{-1/2}e^{x/2}(-x)^{-1/4}n^{-1/4}e^{2\sqrt{-nx}}\left(1+\mathcal{O}\left(n^{-1/2}\right)\right),\ \ \ \ \ \ \ \ x\in\mathbb{C}\backslash\mathbb{R}_{+},
\end{equation}
it is clear that the uniform convergence of the series in \eqref{eq:cf-2} is dominated by the probability $P_{n}^{0}$ whose exponential is linear in $n$.  
  The compact expression of \eqref{eq:cf-2} is given by \cite{Talkner2}
\begin{equation}
{G}(\mu)=\exp\left(\frac{i\mu |f(\tau)|^2}{\hbar\omega}+(e^{i\mu\hbar\omega}-1)|z|^2-4|z|^2\frac{{\rm sin}^2(\hbar\omega\mu/2)}{e^{\beta\hbar\omega}-1}\right).
\label{eq:cf-tilde}
\end{equation}
  Notice that evaluation at $\mu=i\beta$ renders the Jarzynski equality such that $\Delta {F}=|f(\tau)|^2/(\hbar\omega)$.  Also, consistently with the PMWF the minimum of ${G}(2i\beta)$ with respect to the rapidity is obtained at $|z|^2=0$.  For this type of work protocol there is no local minima in ${\rm var}(e^{-\beta {W}_{\rm dis}})$, in contrast with the behavior depicted in Fig.~1 in the main text.  Indeed, using \eqref{eq:cf-tilde}, we find
\begin{equation}
{\rm var}(e^{-\beta {W}_{\rm dis}})=e^{2|z|^2\sinh(\beta\hbar\omega)}-1,
\end{equation}
which does not suffer from divergence at any given $\beta$.

%
\section{Work fluctuations for a driven infinite square-well potential}
%


\begin{figure}[htbp]
   \centering
\includegraphics[width=5in]{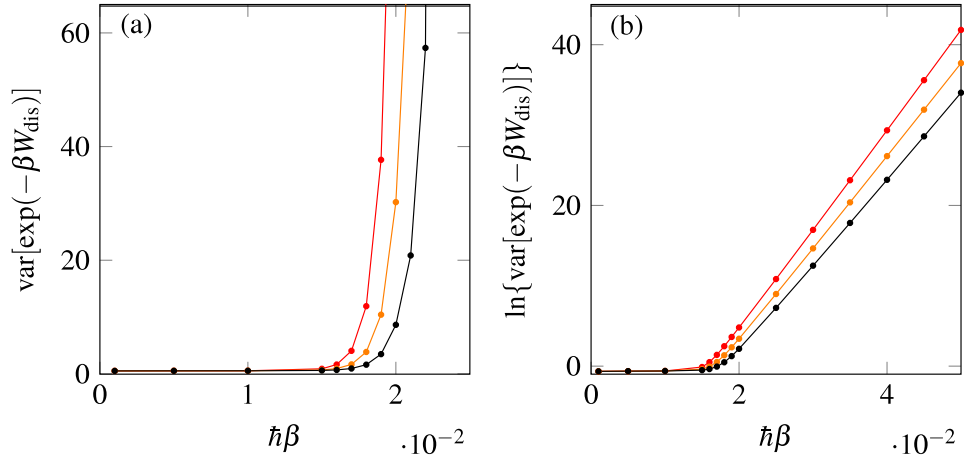}
   \caption{Divergence behavior of the variance of exponential quantum work, for  a compression work protocol applied to particle in a box, with the initial width and final width of the box given by $L_0=2$ and $L_1=1$, with a constant velocity of the moving wall: $v=0.5$.  Numerics with truncation of the double summation in Eq.~(\ref{eq1}) at $\max(n)=\max(m)=\ 32$ (red), 31 (orange), 30 (black).  There is no compact expression for the transition probability $P_{m,n}^{0\rightarrow \tau}$, depicted in \eqref{eq:prob}; as the expression involves an infinite sum that can only be done numerically.  In panel (b) we plot the logarithm of ${\rm var}[\exp(-\beta W_{\rm dis})]$.  It shows a linear trend with different tilts for each truncation.  This indicates clearly an exponential growth in panel (a) with arguments determined by the truncation.  We believe this is a signature of divergence baring resemblance of truncation of a divergent series of the form: $\gamma=\sum_{n=1}^{{\rm max}(n)<\infty} e^{-\epsilon\ n^2}\approx e^{-\epsilon\ {\rm max}(n)^2}$,\ ${\rm log}(\gamma)\approx -\epsilon\ {\rm max}(n)^2$, for some constant $\epsilon<0$.}
   \label{fig:box}
\end{figure}

We present in Fig.~\ref{fig:box} results for an infinite square-well potential, i.e., $H_{box}(t)=-\frac{1}{2M}\hat{p}^2+V_{t}(\hat{q})$, where
\begin{equation}
V_t(q)=\left\{\begin{array}{cc}0,& -L(t)/2<q<L(t)/2,\\ \infty,& {\rm otherwise};\end{array}\right.
\end{equation}
 with a constant velocity of compression, $dL/dt$.  Again, analytical results are not available yet numerics shows the qualitative behavior characteristic of divergence in ${\rm var}[\exp(-\beta W_{\rm dis})]$.  Although not shown in the figures we confirmed the following observations: (i) benchmark for the numerics with the adiabatic divergence, which can be easily predicted at $L_1>\sqrt{2}\ L_0$. (ii) For larger velocities we observe a shift of the divergent behavior towards higher temperatures.  We used the reported transition probability from Ref.~\cite{Quan2}:
\begin{gather}\label{eq:prob}
P_{m,n}^{0\rightarrow \tau}=\left|\sum_{l=1}^{\infty}\ \left\{\frac{2}{L_0}\int_{0}^{L_0} {\rm exp}\left(-i\frac{Mv}{2\hbar L_0}x^2\right){\rm sin}\left(\frac{l\pi x}{L_0}\right){\rm sin}\left(\frac{m\pi x}{L_0}\right)dx\right.\right.\\
\left.\left.\times\ {\rm exp}\left[-il^2\ \frac{\pi^2\hbar(L_1-L_0)}{2MvL_0L_1}\right]\ \frac{2}{L_1} \int_{0}^{L_1} {\rm exp}\left(i\frac{Mv}{2\hbar L_1}y^2\right){\rm sin}\left(\frac{l\pi y}{L_1}\right){\rm sin}\left(\frac{n\pi y}{L_1}\right)dy\right\}\right |^2,\nonumber
\end{gather}
where we set $M=1$ and $\hbar=1$.  The $n$-th energy level of the system being $E^{t}_{n}=n^2\pi^2\hbar^2/(2ML^2(t))$.   Our numerical calculation is based on truncation of the series in Eq.~\eqref{eq1}.
As shown in Fig.~\ref{fig:box}, for work protocols with finite ${\rm var}[\exp(-\beta W_{\rm dis})]$ at high temperatures, divergence will emerge as temperature decreases.  This again confirms our general insights. \\

\end{widetext}

\pagebreak
\bibliographystyle{unsrt}

\end{document}